\documentclass[prl,aps,twocolumn,tightenlines,floatfix,nofootinbib]{revtex4}

\usepackage{epsf}    
\usepackage{graphicx}

\newcommand{\be}{\begin{equation}}
\newcommand{\ee}{\end{equation}}
\newcommand{\bear}{\begin{eqnarray}}
\newcommand{\eear}{\end{eqnarray}}

\newcommand{\rx}{{\rm x}}

\newcommand{\rn}{{\rm n}}
\newcommand{\rp}{{\rm p}}
\newcommand{\rc}{{\rm c}}
\newcommand{\rnp}{{\rm np}}

\begin{document}

\title{A hydrodynamical trigger mechanism for pulsar glitches}

\author{Kostas Glampedakis$^1$ and  Nils Andersson$^2$ }
\affiliation{$^1$SISSA/International School for Advanced Studies and INFN, via Beirut 2-4, 34014 Trieste, Italy \\
$^2$School of Mathematics, University of Southampton, Southampton SO17 1BJ, UK}

\begin{abstract}
The standard explanation for large pulsar glitches involves transfer of angular momentum
from an internal superfluid component to the star's crust. This model requires an instability to 
trigger the sudden 
unpinning of the vortices by means of which the superfluid rotates.
This Letter describes a new instability that may play this role.
The instability, which is associated with the inertial r-modes of a superfluid neutron star, 
sets in once the rotational lag in the system reaches a critical level. 
We demonstrate that our model is in good agreement with observed glitch data, suggesting that 
the superfluid r-mode instability may indeed be the mechanism that triggers large pulsar glitches. 
\end{abstract}

\maketitle


{\em Introduction}.--- Even though pulsars are generally very stable rotators, in some cases with an accuracy that rivals 
the best terrestrial atomic clocks,  many systems exhibit a variety of timing ``noise''. The most enigmatic
features are associated with the so-called glitches, sudden spin-up events followed by a 
relaxation towards steady long-term spin-down. Several hundred glitches, with magnitude
in the range 
$\Delta \Omega_\rc/\Omega_\rc \approx 10^{-9} - 10^{-6}$ 
where $\Omega_\rc $ is the observed rotation frequency, have now been observed in over
100 pulsars \cite{database}. The archetypal glitching pulsar is Vela, which exhibits regular large glitches. 
Glitches have also been reported in several magnetars \cite{axps} as well as one, very slowly rotating, accreting 
neutron star \cite{galloway}. 

Despite the relative wealth of observational data our theoretical understanding of glitches
has not advanced considerably in recent years. 
The standard ``model'' for large pulsar glitches envisages a sudden transfer
of angular momentum from a  superfluid component to the rest of the star \cite{itoh}, which includes 
the crust (to which the pulsar mechanism is assumed rigidly attached) and the charged matter in the core.
A superfluid rotates by forming a 
dense array of vortices, and the vortex configuration determines the global rotation.
The key idea for explaining glitches is that, if the superfluid vortices are pinned to the other component, a rotational lag builds
up as the crust spins down due to electromagnetic braking. Once the rotational lag reaches some critical level, the pinning
breaks. This allows the vortices to move, which leads to a transfer of angular momentum
between the two components and the observed spin-up of the crust.  

Most  theoretical work has focused on either the strength of the vortex pinning \cite{pizzochero,jones} or the 
post-glitch evolution \cite{alpar}. There have not been many suggestions for the mechanism that 
triggers the glitch in the first place. It is generally expected that this role will be
played by some kind of instability, but there are few
truly quantitative models. The results presented in this Letter change the situation dramatically. 
We present evidence for a new  instability, acting on the inertial modes of a rotating superfluid star,
that sets in beyond a critical rotational lag. The predictions of this model agree well with the 
observational data making it plausible that this mechanism provides a missing piece
in the pulsar glitch puzzle.


{\em Inertial mode analysis}. --- We want to improve our understanding 
of the global hydrodynamics associated with a pulsar glitch. Even though this should be a key issue, it has not 
 been discussed in detail previously. In principle, one would expect to be able to express the dynamics in terms of 
global oscillation modes of the system. In this Letter we present the first results for inertial modes of 
neutron star models with the two main features required for the standard glitch models, a superfluid neutron 
component that rotates at a rate different from that of the crust and pinned neutron vortices.
 
We use the standard two-fluid model for superfluid neutron 
stars (see for example \cite{prix}), identifying the two components with the neutron superfluid
and a  conglomerate of all charged particles (the ``protons''). In the following, 
the index $\rx = \{\rn,\rp\} $ identifies the distinct fluids. 
Our aim is to model small amplitude oscillations with respect to a background configuration
where both fluids rotate rigidly with (parallel)
angular velocities $v_\rx^i = \epsilon^{ijk} \Omega_j^\rx  x_k$ and where the magnitudes are different,
$\Omega_\rn \neq \Omega_\rp$.
The linear perturbations of this system
(assuming a time dependence $\sim \exp(i\sigma t) $) are, in the inertial frame, 
described by the two coupled Euler equations, 
\bear
&& \left ( i\sigma + v_\rn^j \nabla_j \right  ) \delta v_\rn^i 
+ \delta v_\rn^j \nabla_j v_\rn^i   +
\nabla^i \delta \psi_\rn  = \delta f_{\rm mf}^i
\label{eulern}
\\
&& \left ( i\sigma + v_\rp^j \nabla_j \right  ) \delta v_\rp^i 
+ \delta v_\rp^j \nabla_j v_\rp^i   +
\nabla^i \delta \psi_\rp  = -\frac{\delta f_{\rm mf}^i}{x_\rp} 
\label{eulerp}
\eear 
Here $\delta \psi_\rx = \delta \tilde{\mu}_\rx + \delta \Phi $ represents the sum of the perturbed 
specific chemical potential and gravitational potential. We have also introduced $x_\rp = \rho_\rp/\rho_\rn $.
This ratio, which is roughly equal to the proton fraction, is assumed constant throughout the star.  
For simplicity, we assume that the two fluids are incompressible, which means that 
$\nabla_i \delta v_\rx^i=0$. In general, the two fluids are coupled
i) chemically, ii) gravitationally,  iii) via the entrainment effect, and iv) by the vortex mediated mutual friction $f_{\rm mf}^i$.  
For clarity, we will ignore the entrainment in the present analysis. A detailed discussion of how the 
entrainment affects our results will be provided elsewhere.

For the inertial modes, the main coupling mechanism is provided by the
mutual friction force. The general expression for this force is \cite{HV},
\be
f_i^{\rm mf} =  {\cal B} \epsilon_{ijk} \epsilon^{kml} \hat{\omega}^j_\rn \omega_m^\rn w^\rnp_l
+  {\cal B}^\prime \epsilon_{ijk} \omega^j_\rn w^k_\rnp
\label{mf}
\ee
where $w^i_\rnp = v^i_\rn -v^i_\rp$ and $\omega^i_\rn = \epsilon^{ijk} \nabla_j v_k^\rn $. 
A ``hat'' denotes a unit vector. When the two fluids are not corotating, the perturbed 
force $\delta f_\mathrm{mf}^i$ is quite complex \cite{sidery08}. 
The form (\ref{mf}) for $f^i_{\rm mf} $ results from balancing the
Magnus force that acts on the neutron vortices and a resistive ``drag'' force which represents the
interaction between the vortices and the charged fluid. Representing the drag force
by a dimensionless coefficient $\cal R$, one finds that 
$\mathcal{B}' = \mathcal{R}\mathcal{B} = \mathcal{R}^2/(1 + {\cal R}^2)$.
The range of values that $\mathcal{R}$ takes in a neutron star is not well known. 
The standard assumption in studies of neutron star oscillations has been that the drag is weak,
which means that $\mathcal{B}'\ll\mathcal{B}\ll 1$. Then the second term in (\ref{mf}) has no effect on the dynamics. 
However, it may well be the opposite limit that applies. The vortices in a neutron star core may  
experience a strong drag force  if their interaction with the 
magnetic fluxtubes is efficient \cite{ruderman,link03}. The drag on the superfluid in the crust may also 
be strong due to vortex ``pinning'' by the lattice nuclei. Even though the current evidence \cite{pizzochero, jones} 
favours weak crustal pinning, the existence of strong pinning regions has not been completely ruled out. 
In these cases one must consider the 
strong coupling limit $ {\cal R} \gg 1 $ which leads to  ${\cal B} \approx 0$ and ${\cal B}^\prime \approx 1 $. 

The hydrodynamical equations (\ref{eulern})-(\ref{eulerp}) allow for a rich set of oscillation modes. 
Our focus will be on a subset of the inertial modes, the purely axial r-modes. The r-modes
have attracted considerable attention since they may suffer a gravitational-driven instability \cite{NAreview}.
They have been studied in superfluid neutron stars previously \cite{lee,prix04,2stream}, but not accounting for both mutual friction
and a rotational lag. When expressed in terms of spherical harmonics, 
the r-mode velocity fields take the form 
\be
\delta v_\rx^i =  \left (  -{im U_l^\rx Y_l^m \over r^2\sin^2 \theta}  
\hat{e}_\theta^i
+{ U_l^\rx \partial_\theta Y_l^m \over r^2 \sin\theta}  \hat{e}_\varphi^i \right ) e^{i\sigma t}
\label{decomp}
\ee
In the corresponding single fluid problem such purely axial solutions exist, corresponding to a single 
$l=m$ multipole. The solution (\ref{decomp}) satisfies the continuity equations automatically, which 
means that we only have to consider the two Euler equations.
Inserting (\ref{decomp}) in (\ref{eulern})-(\ref{eulerp}) and eliminating 
$\delta\psi_\rx $  one arrives at two equations for $U_\ell^\rx $. As in the single fluid 
inertial mode problem, these relations feature couplings between different
$l $ multipoles. In the present problem these couplings are further complicated by the 
presence of $\delta f^i_{\rm mf} $. As a result, one would expect the problem to be difficult to solve. 
Remarkably, this is not the case. It turns out that, even when $\Omega_\rn \neq \Omega_\rp $, 
there exists a simple $l = m $ solution of the form  $U^{\rm x}_m = A_{\rm x} r^{m+1}$.
This solution is exact at leading order in rotation and for all values of 
${\cal R} $. The amplitudes $A_\rx$ follow from the solution of
a $2\times2 $ algebraic system, the determinant of which provides the dispersion
relation for the mode frequency. 

{\em A new superfluid instability}.---
The strong coupling limit, ${\cal B} =0$ and ${\cal B}^\prime =1 $, provides a good illustration 
of the main new result. In this case we obtain a simple mode solution.
In terms of the dimensionless parameters 
\be
\kappa = \frac{\sigma + m\Omega_\rp}{\Omega_\rp}, \quad 
\Delta = \frac{\Omega_\rn -\Omega_\rp}{\Omega_\rp}
\ee
we have the two frequency solutions
\be
\kappa_{1,2} = -\frac{1}{(m+1) x_\rp} \left [ 1 -x_\rp + \Delta \pm {\cal D}^{1/2} \right ]
\label{strongroots}
\ee 
where
\be
{\cal D} = (1+x_\rp)^2 + 2\Delta \left [ 1 + x_\rp \left \{ 3 -m(m+1) \right \} \right ] + O(\Delta^2)
\ee
The amplitudes are related according to
\be
\frac{A_\rn}{A_\rp} = \frac{2\left (1 + \Delta  \right )}{(m+1)\kappa}
\label{amplis}\ee

Let us focus on the short-lengthscale modes. Taking $m\gg 1$ and recalling that $x_\rp$ and $\Delta$ are both 
small (generally $\Delta \ll x_\rp$), it is easy to show that
one of the above  r-mode solutions becomes unstable
(${\rm Im}[\kappa] <0 $) for  $m > m_c$, where 
\be
m_c \approx \frac{1}{\sqrt{2x_\rp \Delta}}
\approx 320 \left ( \frac{0.05}{x_\rp} \right )^{1/2} \left ( \frac{10^{-4}}{\Delta} \right )^{1/2}
\label{mc}
\ee
For $m \gg m_c$ the instability growth timescale $\tau_{\rm grow} = 1/(\Omega_\rp{\rm Im}[\kappa])$ is
well approximated by 
\be
\tau_{\rm grow} \approx \frac{P}{2\pi}  \sqrt{ \frac{x_\rp}{2\Delta}}  
\approx 0.25 \left ( \frac{x_\rp}{0.05} \right )^{1/2}
\left ( \frac{10^{-4}}{\Delta} \right )^{1/2} \left( { P \over 0.1\ \mathrm{s}}\right)\ \mathrm{s}\, 
\label{grow}
\ee
where $ P = 2\pi/\Omega_\rp $ is the observed spin period 
(we associate the charged component with the crust, $\Omega_\rp = \Omega_\mathrm{c}$). We see that this new instability
can grow  rapidly, on a timescale comparable to the rotation period of the star.

Although we cannot yet claim to understand the detailed nature of this new r-mode instability, we have some
useful clues. Firstly, we find from (\ref{amplis}) that the unstable modes are such that
$|A_\rn/A_\rp| \sim x_\rp$. Thus, the fluid motion is predominantly in the proton fluid. 
There should also be a close connection with the short wavelength instability that we recently discovered 
for precessing superfluid stars \cite{preclett}. In that case, the result followed from a local plane-wave analysis 
of the inertial modes. An attempt to link these two results 
would be useful. Finally, since the present system has two distinct rotation rates
one might expect the instability to belong to the general two-stream class \cite{2stream}.
Such instabilities are generic in multifluid systems. The
intuitive condition for such an instability dictates that the mode's pattern speed $-{\rm Re}(\sigma)/m $ 
should lie between $\Omega_\rn $ and $\Omega_\rp $ \cite{2stream} . This criterion translates into 
$m>\sqrt{2} m_c$, which is satisfied in the main part of the parameter space where the
instability is operative. 

In order for the new instability to affect the dynamics of
realistic neutron stars it must (at least) overcome viscous damping.  
For young and mature neutron stars, dissipation is dominated by shear viscosity due
to electron-electron collisions \cite{FI76}. For a uniform density star with  
$M=1.4M_\odot$ and $R=10$~km (the canonical values used in the following) 
the corresponding  viscosity coefficient is \cite{viscous} 
\be
\eta_\mathrm{ee} \approx 2.7 \times  10^{20} \left ( \frac{x_\rp}{0.05} \right )^{3/2}
\left ( \frac{10^8 \mathrm{K}}{T} \right )^{2}\, {\rm g}\,{\rm cm}^{-1}\, {\rm  s}^{-1} 
\ee
where $T$ is the core temperature. We can use the standard energy-integral approach
to estimate the viscous damping timescale. In fact, since $|A_\rn| \ll |A_\rp|$ we can use existing results for r-modes of 
uniform density stars \cite{kokk}, remembering that shear viscosity only acts on the proton fluid. 
Thus, a simple calculation leads to 
\be
\tau_\mathrm{sv} \approx \frac{ 3 M x_\rp}{ 8 \pi m^2 \eta_{\rm ee} R} 
\approx \frac{6\times10^4}{m^2} \left ( \frac{0.05}{x_\rp} \right )^{1/2}
\left ( \frac{T}{10^8\, {\rm K}} \right )^2\, {\rm s} 
\label{tau_sv}
\ee
Combining (\ref{grow}) and (\ref{tau_sv}) we have a criterion for the 
unstable modes to grow fast enough to overcome viscous damping. The condition 
$\tau_\mathrm{grow} < \tau_\mathrm{sv}$ leads to
\be
m < 500 \left (\frac{0.05}{x_\rp} \right )^{1/2} \left ( \frac{\Delta}{10^{-4}} \right )^{1/4}
\left (\frac{T}{10^8\,{\rm K}} \right ) \left ( \frac{0.1\,{\rm s}}{P} \right )^{1/2}
\ee
This gives the range of unstable $m$ modes.  Since the instability sets in 
when $m > m_c$ we conclude that the system becomes unstable
once it reaches the critical lag
\be
\Delta_c \approx 6 \times 10^{-5}  \left( { P \over 0.1\ \mathrm{s} } \right)^{2/3}
\left( { T \over 10^8\ \mathrm{K} } \right)^{-4/3}
\label{Delc}\ee 


{\em Making contact with observations}.--- 
Within a two-component model, it is straightforward to estimate the critical lag required to explain the observations.
Assuming that angular momentum is conserved in the process, one must have 
$I_\rc \Delta \Omega_\rc \approx - I_{\rm s} \Delta \Omega_\mathrm{s}$, 
where $I_{\rm s}$ and $I_{\rm c} $ are the two moments of inertia, while  
$\Delta \Omega_\mathrm{s}$ and $\Delta \Omega_\mathrm{c}$ represent the changes in the corresponding spin frequencies. 
The glitch data suggests that about $2\,\%$ of the total spin-down is 
reversed in the glitches \cite{lyne00}, suggesting that $I_{\rm s}/I_{\rm c}\approx 0.02$. 
In order to permit Vela-sized glitches with $\Delta \Omega_\rc/\Omega_\rc \sim 10^{-6}$ we then need 
(assuming $\Omega_c = \Omega_s $ after the event)
\be
\Delta_g \approx {I_\rc \over I_{\rm s} }{ \Delta \Omega_\rc \over \Omega_\rc} \approx  5\times 10^{-4}
\label{Del}
\ee

The observational estimate of the lag $\Delta_g $ at which large glitches occur is  close to our estimate 
(\ref{Delc}) for the onset of the superfluid r-mode instability. This is unlikely to be a coincidence. 
Even though it is difficult to compare  the
parameters of our two-fluid neutron star model to the global quantities used in the 
phenomenological discussion directly, it is clear that our new instability 
has the features expected of a glitch trigger mechanism. 
It operates in the strong drag limit, where vortices are effectively pinned to the charged component.
As long as the system is stable, a rotational lag should build up as the crust spins down.   
Once the system evolves beyond the critical level (\ref{Delc}) a range of unstable r-modes grow on a timescale 
of a few rotation periods. We cannot yet say what happens when these modes reach large 
(non-linear) amplitudes, but it seems inevitable that the fluid
motion associated with the instability will break the vortex pinning, allowing a glitch to proceed.

\begin{figure}
\centerline{\includegraphics[height=6cm,clip]{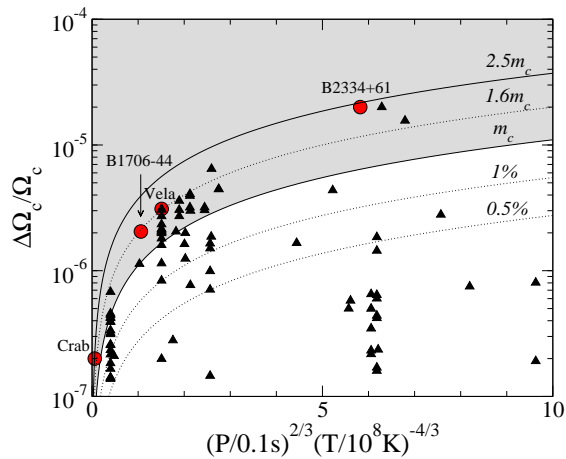}}
\caption{The maximum glitches predicted from (\ref{Delc}) are compared to 
observations. The $m_c$ curve 
represents the onset of the superfluid instability (the instability region is grey, and for $2.5m_c$ there is 
a range of unstable inertial modes) and a glitch involving
the relaxation of 2\% of the total moment of inertia (corresponding curves for 1\% and 0.5\% are also shown). 
The circles are systems with temperature data  (taken from \cite{pons}). The triangles represent 
systems for which $T$ is estimated using (\ref{coolfit}).  The glitch-data is taken from 
\cite{database,middle}. }
\label{critplot}
\end{figure}

Let us compare the ``predictions'' of our model to the data for pulsars exhibiting large glitches. 
To do this, we estimate the  maximum glitch size allowed if $\Delta_g =\Delta_c$, assuming a completely
relaxed system and  $I_\mathrm{s}/I_\rc=0.02$. 
Since we do not have temperature data for most 
glitching pulsars we estimate $T$ by combining the heat blanket model from \cite{gudmundsson}
with a simple modified 
URCA cooling law. Calibrating this model to the Vela pulsar, for which $T\approx 6.9\times 10^7$~K \cite{pons}, we find
\be
T \approx 3.3 \times 10^8\ \left ( { t_c \over 1 \ \mathrm{yr}} \right)^{-1/6}\ \mathrm{K}
\label{coolfit}
\ee
Here $t_c=P/2\dot{P}$ is the characteristic pulsar age. The results are shown 
in Fig.~\ref{critplot}. This Figure shows that our model does well in predicting the
maximum glitches one should expect. The data is consistent with the idea that a
system needs to evolve into the instability region before a large glitch happens. 
It should be noted that, even though the instability first appears at $m=m_c$, the growth time is much longer 
than (\ref{grow}) until $m>1.2m_c$ or so. It is also interesting to note that two of the systems with
actual temperature data \cite{pons}, Vela and PSR  B1706-44, both sit on the
 $m\approx 1.6 m_c$  curve.

{\em Discussion}. --- We have described a new instability that may operate in rotating superfluid neutron stars. 
We have demonstrated that this instability sets in at parameter values that compare well 
with those inferred from pulsar glitches. This suggests that the superfluid
r-mode instability may be the mechanism that triggers large pulsar glitches.

The model is consistent with a number of observed properties of glitching pulsars:

\noindent
i) Adolescent pulsars, like Crab and PSR J0537-69,  should only 
exhibit small amplitude glitches. For fast spin and a relatively high temperature 
the instability sets in at smaller values of  $\Delta $.

\noindent
ii) More mature and slower spinning neutron stars have colder cores which
means they can produce larger glitches,  provided the
required $\Delta$  can build up (cf. PSR J1806-21
with $\Delta\Omega_\rc/\Omega_\rc \approx 1.6 \times 10^{-5} $ \cite{hobbs}). 
Since $\Delta_c$ increases
as the star ages one would expect neutron stars to cease to glitch eventually.

\noindent
iii) For any glitch mechanism that relies on a critical spin lag
between a superfluid component and the rest of the star it is easy to 
estimate the time interval $t_g $ between successive glitches. Assuming that each 
glitch relaxes the system, we estimate 
$t_g \approx 2 \Delta_g  t_c \gtrsim 2 \Delta_c t_c $. For Vela we then
find $t_g \gtrsim 750\,{\rm d}$ which compares well with the
observed averaged time of about $1000\,{\rm d}$. The most regular known glitcher, 
PSR J0537-69, has an average interglitch time of about $120\,{\rm d}$ \cite{middle}.
In the absence of temperature data we use (\ref{coolfit}) for this object   
and find $ t_g \gtrsim 90\,{\rm d}$. Again, the agreement with the observations is 
good, and consistent with the notion that the system evolves into the unstable regime before a glitch occurs.

\noindent
iv) There is no reason why the instability should not 
operate in all spinning neutron stars in which a rotational lag builds up. In particular, one may expect
accreting neutron stars to ``glitch'' occasionally. So far, there has only been one suggested event, in 
the slowly rotating transient KS 1947+300 \cite{galloway}. For this system our model suggests  (combining $P=18.7$~s
with $T\approx 10^8$~K, a temperature that should be typical of an accreting star) a 
maximum glitch of  $\Delta \Omega_\rc / \Omega_\rc \approx 4\times 10^{-5}$. This is very close to the suggested 
observed glitch with $\Delta \Omega_\rc / \Omega_\rc \approx 3.7\times 10^{-5}$ \cite{galloway}. 
Such events should, of course, be extremely rare. 

Since we have considered the non-magnetic inertial mode problem, our 
model  does  not apply (without modification) to magnetars. 
It is nevertheless interesting to consider these systems. Given typical 
magnetar parameters ($P\sim 10\,{\rm s}$, 
$ T \sim 10^9\,{\rm K} $), we would not expect these objects to exhibit
large glitches. Yet, they do \cite{axps}.  The resolution may be that these glitches 
involve a larger fraction $I_\mathrm{s}/I_\mathrm{c}$.     
A very interesting question for the future concerns whether this is  a natural 
consequence of stronger magnetic pinning in the neutron star core.   

These first results are promising, but we are obviously far away from a complete
understanding of this new mechanism. Future work needs to consider more 
detailed neutron star models. We need to understand the local mutual friction parameters 
and the nature of  vortex pinning. The inertial mode problem for realistic neutron star models presents a 
real challenge. We should also make more detailed attempts at understanding the observations. A natural 
first step would be to  improve on the temperature estimates used in Figure~\ref{critplot}. 
Finally, we need to understand the nonlinear development of the new instability.
This problem can perhaps be studied with numerical simulations,  building on the work discussed in 
\cite{peralta}. Ultimately one would hope to arrive at a truly quantitative model 
for pulsar glitches.




\begin{thebibliography}{10}

\bibitem{database}
www.atnf.csiro.au/research/pulsar/psrcat/

\bibitem{axps}
R. Dib, V.M. Kaspi and F.T. Gavriil, Ap. J. {\bf 673}, 1044 (2008)

\bibitem{galloway}
D.K. Galloway, E.H. Morgan and A.M. Levine, Ap. J. {\bf 613}, 1164 (2004)


\bibitem{itoh}
P.W. Anderson and N. Itoh, Nature {\bf 256 }, 25 (1975) 

\bibitem{pizzochero}
P.M. Pizzochero, L. Viverit and R.A. Broglia, Phys. Rev. Lett. {\bf 79}, 3347 (1997)  

\bibitem{jones}
P.B. Jones, MNRAS {\bf 296 }, 217 (1998)

\bibitem{alpar}
M.A. Alpar {\em  et al.}, Ap.\, J.\, {\bf 276}, 325
(1984) 


\bibitem{prix}
R. Prix, Phys. Rev. D {\bf 69} 043001 (2004)



\bibitem{HV}
H.E. Hall and W.F. Vinen, Proc. R. Soc. A {\bf 238}, 215 (1956)

\bibitem{sidery08}
T. Sidery, N. Andersson and G.L. Comer, MNRAS {\bf 385}, 335  (2008)



\bibitem{ruderman}
M. Ruderman, T. Zhu and K. Chen, Ap.\,J.\, {\bf 492},  267 (1998)

\bibitem{link03}
B. Link, Phys. Rev. Lett., {\bf 91},  101101 (2003) 


\bibitem{NAreview}
N. Andersson, Class. Quantum Grav. {\bf 20}, R105 (2003)



\bibitem{lee}
U. Lee and S. Yoshida, Ap. J. {\bf 586}, 403 (2003) 


\bibitem{prix04}
R. Prix, G.L. Comer and N. Andersson, MNRAS {\bf 348}, 625 (2004)


\bibitem{2stream}
N. Andersson, G.L. Comer and R. Prix, MNRAS {\bf 354}, 101 (2004)


\bibitem{preclett}
K. Glampedakis, N. Andersson and D.I. Jones, Phys. Rev. Lett. {\bf 100}, 081101 (2008)


\bibitem{FI76}
E. Flowers and N. Itoh, Ap. J. {\bf 206}, 218 (1976)


\bibitem{viscous}
N. Andersson, G.L. Comer and K. Glampedakis, Nucl. Phys. A {\bf 763}, 212 (2005)


\bibitem{kokk}
K. D. Kokkotas and N. Stergioulas, A\&A {\bf 341}, 110 (1999)


\bibitem{lyne00}

A.G. Lyne, S.L. Shemar and F. Graham Smith, MNRAS {\bf 315}, 534 (2000)


\bibitem{gudmundsson}
E.H. Gudmundsson, C.J. Pethick and R.I. Epstein, Ap.\,J.\, {\bf 259}, L19 (1982) 

\bibitem{pons}
D. N. Aguilera, J.A Pons and J.A. Miralles, Ap.\, J.\, {\bf 673}, L167 (2008)

\bibitem{hobbs}
G. Hobbs {\it et al.}, MNRAS {\bf 333 }, L7 (2002) 



\bibitem{middle}
J. Middleditch {\it et al.}, Ap.\, J.\, {\bf 652}, 1531 (2006)


\bibitem{peralta} 
C. Peralta {\em et al.}, Ap.\, J.\, {\bf 635} 1224 (2005); {\it ibid}. {\bf 651}, 1079 (2006)


\end{thebibliography}
\end{document}